\documentclass[figures,preprint,cite,bm]{epl}
\usepackage{amsmath}
\usepackage{amssymb}

\newcommand{\strainepsilon}{\varepsilon}

\title{
Continuum mesoscale theory inspired by plasticity
}
\date{\today}

\author{James P. Sethna \inst{1} 
\thanks{E-mail: \email{sethna@lassp.cornell.edu},
Home page: \email{http://www.lassp.cornell.edu/sethna/sethna.html}} 
\and Markus Rauscher \inst{1,2,3}
\thanks{E-mail: \email{rauscher@mf.mpg.de}}
\and Jean-Philippe Bouchaud \inst{4}
\thanks{E-mail: \email{bouchau@saclay.cea.fr}}
}
\institute{
	\inst{1} Laboratory of Atomic and Solid State Physics (LASSP) - Clark
	Hall, Cornell University, Ithaca, NY 14853-2501, USA\\
	\inst{2} Max-Planck-Institute f\"{u}r Metallforschung - Heisenbergstr.\
	3, 70569 Stuttgart, Germany\\
	\inst{3} Institut f\"{u}r Theoretische und
	Angewandte Physik - Universit\"{a}t Stuttgart, 70550 Stuttgart,
	Germany\\
	\inst{4} Service de Physique de l'Etat Condens\'e - CEA Saclay, 91191 Gif
	sur Yvette, France
}


\begin{document}
\maketitle

\begin{abstract}
We present a simple mesoscale field theory inspired by rate-independent 
plasticity that reflects the symmetry of the deformation process. We
parameterize the plastic deformation by a scalar field which evolves with
loading. The evolution equation for that field has the form of a
Hamilton-Jacobi equation which gives rise to cusp-singularity
formation. These cusps introduce irreversibilities analogous to those seen
in plastic deformation of real materials: we observe a yield stress, work
hardening, reversibility under unloading, and cell boundary formation.
\end{abstract}

\begin{figure}
\onefigure[width=0.97\linewidth]{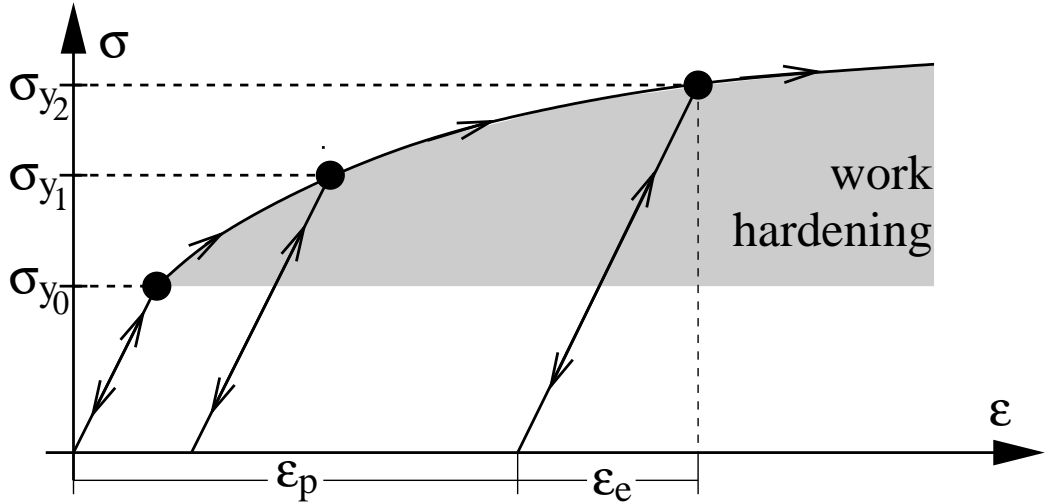}
\caption{\label{realstressstrainfig} Generic stress-strain relation. For
stress $\sigma$ below the critical yield stress ${\sigma_{\text{y}}}_0$ the 
deformation is elastic and reversible. Further deformation results in
plastic deformation and work hardening, \textit{i.e.\/} the yield point
(bullet) moves to higher stress. The strain $\strainepsilon$ is partly due
to reversible elastic deformation $\strainepsilon_{\text{e}}$ and partly
due to irreversible plastic deformation $\strainepsilon_{\text{p}}$.}
\end{figure}

We call it plasticity when materials yield irreversibly at large external
stresses. Macroscopically, plasticity is associated with three qualitative
phenomena. To a good approximation, there is a threshold called the
\textit{yield stress} below which the deformation is reversible (see
Fig.~\ref{realstressstrainfig})\/. A material pushed beyond its yield
stress exhibits \textit{work hardening}, through which the yield stress
increases to match the maximum applied stress.  Finally, the deformed
crystal develops patterns, such as the \textit{cell structures} observed in
fcc metals \cite{kuhlmannwilsdorf89,argon,hughes4}.  While much is known
about all three phenomena, 
a quantitative understanding based on mesoscopic continuum theory would be
welcome---especially if it connects to microscopic properties of the atomic
interactions in the material. In this paper, we will discuss a simple
scalar field theory which naturally exhibits these three key
features of plasticity.
We do not claim to model in details plasticity in real materials but we
believe that a continuum description of plasticity should share the key
feature of our model equation: The transition between reversible and 
irreversible deformation is generated by singularities which 
occur at finite stress.

There are a vast number of approaches to modeling plastic flow in 
metals---quantum, atomistic, motion of single dislocations, motion of many
dislocations, continuum theories of dislocation densities, slip system
theories, work hardening theories \dots\/. Reviewing even only the basic ideas
of these models is beyond the scope of this paper. Three of them, however, directly inspired our approach. (1)~Theories
based on the differential geometry of the Burgers vector density torsion
tensor \cite{kroner80,rickman97}. These elegant mathematical descriptions
of the state of the material need to be supplemented by a similarly
sophisticated dynamical evolution law---especially for the
non-equilibrium problem of plastic flow. The torsion tensor theories are
typically dismissed by the engineering community because they ignore the
large majority of geometrically unnecessary dislocations which cancel
out in the macroscopic Burgers vector density. (2)~Macroscale
engineering theories of plasticity, and in particular the recent strain
gradient theories \cite{fleck93}. Their use of symmetry to
constrain the form of the evolution laws is echoed in our approach.
Their explicit incorporation of a yield surface is clearly appropriate
on the macroscale, but is unsatisfactory for a condensed-matter
physicist. In our model, the yield surface emerges naturally from the
evolution laws through the development of cell structures on the
mesoscale. (3)~Theories which attempt to coarse-grain the complex
rearrangement dynamics of atoms \cite{falklanger,langer01} or
dislocations \cite{groma00} to develop continuum plasticity theories. 
Any such description will introduce a field which
describes the local state of the material as it evolves during deformation.
Even though the microscopic dynamics of dislocations is understood to a
large extent, coarse-graining the microscopic dynamics has not been
successful up to now. 

We are going to study the following evolution equation for a scalar field
which parameterizes rate-independent plastic deformation
\begin{equation}
\label{MostGeneralEquation}
\frac{\partial \Psi}{\partial t} = 
\frac{\partial S_{ij}}{\partial t}\,\nabla_i\Psi\,\nabla_j\Psi.
\end{equation}
We use Einstein's convention, summing over repeated indices.
The plastic deformation is assumed not to depend on volume changes and the
evolution depends only on the deviatoric stress
$S_{ij}=\sigma_{ij}-\frac{1}{3}\,\sigma_{kk}\,\delta_{ij}$,
with $\sigma_{ij}$ the full stress tensor. There are two
motivations to study this equation in the context of plasticity theory.

First, we can derive this evolution equation as one of the leading terms in
a long-wavelength expansion of a general local evolution equation for
$\Psi$ in which respects the symmetries and the rate invariance of the loading process. However,
there is {\it a priori} a term of the same order $\frac{\partial S_{ij}}{\partial t}\,\nabla_i\nabla_j\Psi$ which turns out to be highly 
singular since it it leads to finite time divergences of the field $\Psi$.
Although rate independent (creep) terms such as $\nabla^2 \Psi$ (and higher order derivatives) should regularize these divergencies, we prefer to 
discard these singular terms altogether in the present discussion and 
focus on eq.~(\ref{MostGeneralEquation}) which, as we now show, already 
contains an rich phenomenology\/. We will indeed show later that eq.~(\ref{MostGeneralEquation}) can be
transformed into a Hamilton-Jacobi equation which is known to form finite
time singularities. These singularities are the second reason to study
this equation since they can be related to the onset of irreversibility and
to dislocation structure formation. Second,
Hamilton-Jacobi equations are also
related to conservation laws as studied in the context of traffic-jams and
at least on a naive level there is a similarity between traffic-jams and
dislocation entanglement. 

We will focus on so-called
proportional loading paths (such as those used in shear and tension tests,
where the loading direction does not change)
where the stress $S_{ij}(t)=\sigma(t)\,\hat{S}_{ij}$\/. Thus in
eq.~(\ref{MostGeneralEquation}) we 
change variables from time $t$ to stress amplitude $\sigma$
\begin{equation}
\label{ProportionalLoadingEquation}
\frac{\partial \Psi}{\partial \sigma} = 
\hat{S}_{ij}\,\nabla_i\Psi\,\nabla_j\Psi.
\end{equation}

In the form of eq.~(\ref{ProportionalLoadingEquation}), we see the key
challenge in formulating laws of rate-independent plasticity: the equations
appear to be manifestly reversible.  Increasing and then reducing $\sigma$
will naively leave the material in the original state.  The engineers
bypass this
problem by formulating their theories not in terms of order parameter
fields, but directly in terms of a yield surface (corresponding to step
functions $\Theta(\sigma-\sigma_y)$ in the equations of motion)\/.
However, eq.~(\ref{ProportionalLoadingEquation}) is a Hamilton-Jacobi
equation (closely related to the multidimensional anisotropic Burgers
equation, see \cite{frisch02}) which develops cusp-shaped singularities at
finite $\sigma$ even for smooth initial conditions. 
At these singularities, as numerically illustrated below,
information and thus reversibility is lost. The stress at which the
first singularities form has many features of a yield surface as we will
show later.

After the formation of singularities the solution of
eq.~(\ref{ProportionalLoadingEquation}) is not unique and one has to find a
weak solution which regularizes the singularities. As mentioned above, the assumption of a
rate-independent behavior discards creep relaxation effects which enter
eq.~(\ref{MostGeneralEquation}) in form of a diffusion term
$\nabla_k\nabla_k\Psi$\/ with a rate independent coefficient. Such a term with an infinitesimally small
prefactor is enough to regularizes the singularities and one gets the so-called
viscosity solution, which we will consider in the following.

We solve eq.~(\ref{ProportionalLoadingEquation}) numerically on a finite
difference grid using the problem solving environment
\mbox{\textsc{cactus~4.0}} \cite{cactus}\/.  The initial condition was a
Gaussian random field with amplitude and correlation length one. The
convolution was performed in Fourier space using the fast Fourier transform
package {\textsc{fftw~2.1.3}} \cite{fftw}\/. We use an essentially
non-oscillatory (ENO) scheme in combination with Godunov's method to
minimize numerical damping \cite{osher91}\/. The algorithm is proven to
converge to the viscosity solution. The numerical grid has periodic
boundary conditions; in one dimension (1D) the length of the system is 100
and the system had 4084 points; in two dimensions (2D) the size of the
system is $25^2$ and the grid was $1016^2$\/.  The three-dimensional (3D)
simulation cell was $12.5^3$ and the grid size $264^3$\/.  The ENO stencil
width was $4$ in one and $5$ in 2D and 3D, respectively, and the time
stepping scheme is a simple explicit Euler scheme with a time step
$\delta\sigma=10^{-3}$. The system is cycled by unloading beginning at
various points $\sigma_n$ and reloading at somewhat above zero stress (to
reduce the effect of numerical damping)\/. 

\begin{figure}
\onefigure[width=\linewidth]{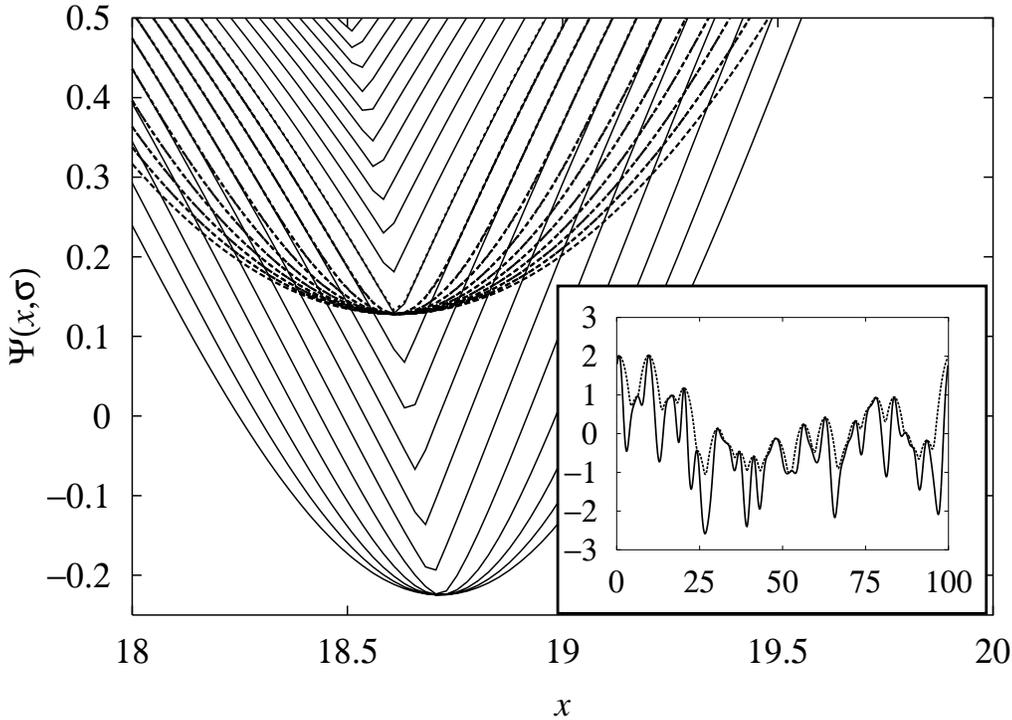}
\caption{\label{cuspfieldin1Dfig} Blowup of the region $18 < x < 20$
from our 1D simulation. The solid lines represent $\Psi(x)$ for $\sigma$
values $0.1$ apart starting at $\sigma=0$, with curves further up having
larger $\sigma$\/.  The dashed lines are $\Psi(x)$ while unloading and the
dotted ones while reloading. The dotted lines are hardly visible because in
the reversible region they lie on top of the dashed ones.  At the cusps,
the spatial discretization is visible.  The inset shows the initial (solid)
and final (dashed) configuration of the whole system, \textit{i.e.\/} for
$\sigma=0$ and $6$\/.}
\end{figure}

Fig.~\ref{cuspfieldin1Dfig} shows the 
evolution of the Burgers equation in 1D, corresponding
to eq.~(\ref{ProportionalLoadingEquation}) with $S_{ij} = 1$\/.
The cusps develop in the valleys at $\sigma\approx
0.3$, and the unloading and reloading paths are both shown. Since
the cusps disappear immediately upon unloading, the order parameter
evolution is reversible until the stress grows to match the previous
maximum, so our model exhibits work hardening. 

\begin{figure}
\twoimages[width=0.47\linewidth]{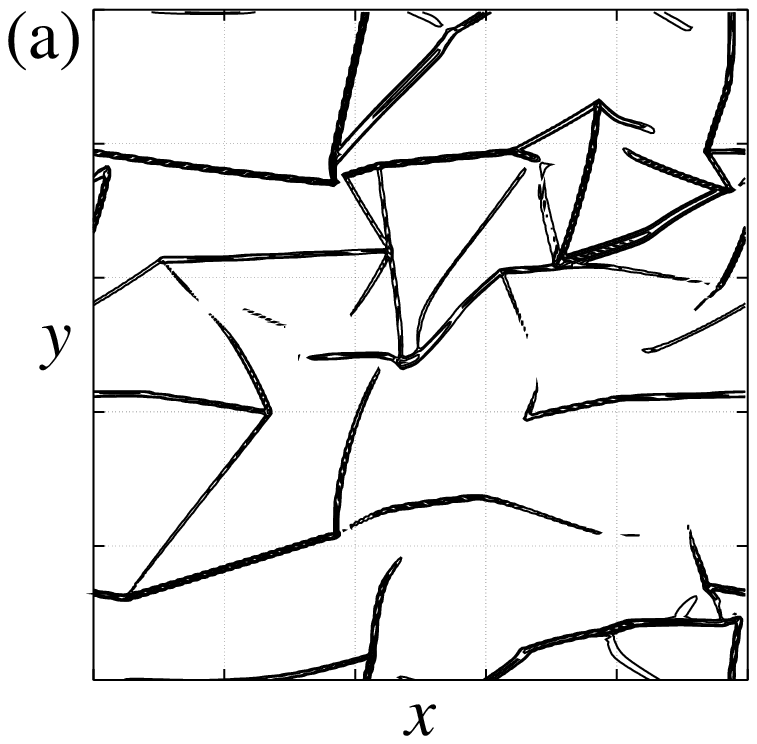}{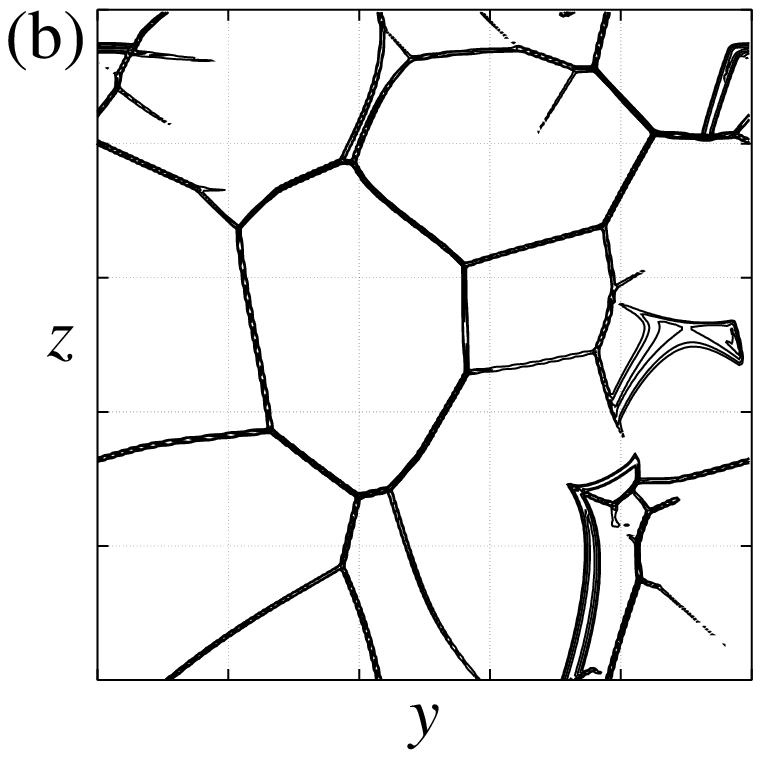}
\caption{\label{burgers3Dshocksfig} Location of the singularities for
tension test at $\sigma=1$ in (a) the $xy$-plane and (b) $yz$-plane, that is 
parallel and perpendicular to the loading direction, respectively. 
The lines are contour lines $\hat{S}_{ij}\nabla_i\nabla_j\Psi = 1.5$, $3$, 
$6$, $9$, \dots\/. 
Our symmetry analysis does not provide us a physical
interpretation for $\Psi$.
}
\end{figure}

Finally, Fig.~\ref{burgers3Dshocksfig} shows the cusp locations in
two cross-sections of a 3D simulation with 
$\hat{S}_{xx}=2/3$, $\hat{S}_{yy}=\hat{S}_{zz}=-1/3$, 
and $\hat{S}_{ij}$ zero otherwise, appropriate to a tension test.
We see that the cusps form
nearly flat 1D interfaces separating cell-like volumes. 
While the cut parallel to the loading direction~(a) shows the expected 
asymmetry, the cut perpendicular to the loading direction~(b)
shows an isotropic hexagonal cell structure.
Simulations under shear in 2D ($\hat{S}_{xx}=1/2$, 
$\hat{S}_{yy}=-1/2$, and zero otherwise) show
similar cusps to Fig.~\ref{burgers3Dshocksfig}(a)\/.
These morphologies are reminiscent of cell structures formed in 
hardened fcc metals \cite{hughes4}. In particular the orientation of the
cell walls is the same as observed in experiments,
because the symmetry if the loading process is reflected in
eq.~(\ref{MostGeneralEquation})\/. This, however, is only
true for isotropic initial conditions.

The size distribution of the cells depends on the initial conditions and
the cells coarsen in contrast to the cell refinement observed in
experiments. Moreover, the cell walls in our model blur upon unloading,
(see Fig.~\ref{cuspfieldin1Dfig}\/), much more than is seen experimentally.
It is clear that these effects call for a more realistic extension of the present model, but we believe that the emergence of {\it spatial structures},
coupled with finite stress singularities which lead to irreversibility, is a generic feature which will be shared by a more elaborated non-linear field 
description of plasticity. 

We now wish to calculate a scalar, \textit{e.g.}, the total plastic strain,
from the $\Psi$ field which illustrates the transition from reversible to
irreversible behavior in a similar way as the stress-strain relation in
Fig.~(\ref{realstressstrainfig})\/. The term we are going to calculate and
which we will call strain is
\begin{equation}
\label{StrainLaw}
\frac{\partial \strainepsilon}{\partial \sigma} = 
\sigma\,\hat{S}_{k\ell}\,\left\langle 
(\nabla_{\!k} \Psi)\, (\nabla_{\!\ell} \Psi) \,
\bm{\nabla}^2\Psi \right\rangle.
\end{equation}
The angle brackets are spacial averages. 
There are two reasons for focusing on this term. 

First it is sensitive to the formation of cusp singularities. At the cusp,
the second derivative gives a delta function. Hence, the appearance of
singularities will greatly affect the strain. 

Second, the term in eq.~(\ref{StrainLaw}) is part of
a gradient expansion. We assume (as common among engineers) that the
deformation is in the direction of the applied stress $S_{ij}$. If we
expand the strain rate $\partial \strainepsilon/\partial t$ to second order
in $S_{ij}$ and fourth order in gradients the term in eq.~(\ref{StrainLaw})
is one of the terms. Other terms are either insensitive to the
singularities or they are surface terms (which would give zero average strain
since we use periodic boundary conditions)\/. There is, however, one extra term
in such an expansion, which is highly singular, namely
$S_{ij}\, \frac{\partial S_{k\ell}}{\partial t}\,(\nabla_{\!k}
\nabla_{\!\ell} \Psi) \, \bm{\nabla}^2\Psi$\/. This term would lead to a
delta-function squared at a cusp, which we again neglect.

\begin{figure}
\onefigure[width=\linewidth]{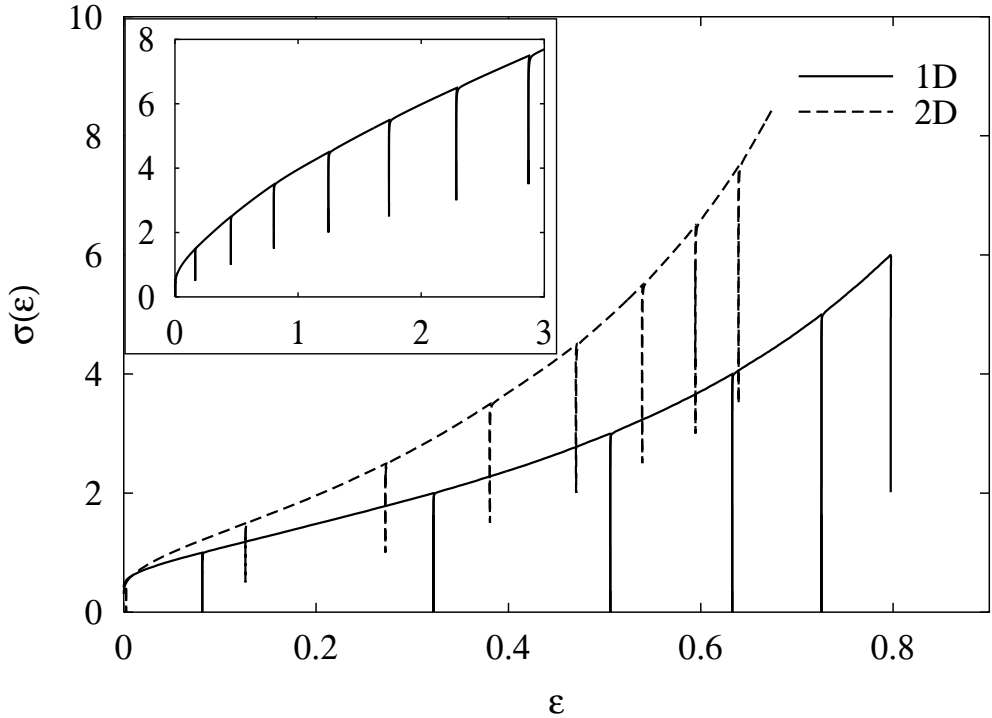}
\caption{\label{burgers12Dstressstrainfig} Stress-strain relation according
to eq.~(\protect\ref{StrainLaw}) for our 1D (solid line) and
2D (dashed line) model, and for our 2D model with a stress-dependent
prefactor $1+0.25\,S_{mn}\,S_{mn}$ (inset)\/.}
\end{figure}

Fig.~\ref{burgers12Dstressstrainfig} 
shows the resulting stress-strain curves for our plasticity theory.  In 1D
and 2D the term in angle brackets in eq.~(\ref{StrainLaw}) is a total
divergence: it is proportional to $\nabla_x (\nabla_x \Psi)^3$ and
$\nabla_x[(\nabla_x\Psi)^3/3-(\nabla_x\Psi)(\nabla_y\Psi)^2] -
\nabla_y[(\nabla_y\Psi)^3/3-(\nabla_y\Psi)(\nabla_x\Psi)^2]$ in 1D and 2D,
respectively.  This implies that $\left\langle {\partial
\strainepsilon_{ij}}/{\partial t}\right\rangle = 0$,
explaining why the unloading curves are
vertical in Fig.~\ref{burgers12Dstressstrainfig}\/. Plastic deformation
occurs on the main loading curve because the cusps act as sources for the
total divergence term. (We do not show stress-strain curves for the 3D
simulations. Because of numerical limitations the results would be
governed by finite size effects, \textit{i.e.}, too few cells in the sample
volume. Qualitatively the stress-strain curves look similar.)

In the inset of Fig.~\ref{burgers12Dstressstrainfig} we added a stress
dependent prefactor to eq.~(\ref{StrainLaw}) in order to change the
curvature of the stress-strain curve and make it look more similar to
Fig.~(\ref{realstressstrainfig})\/. This is to demonstrate that the
upward-bending of the stress-strain curve is more related to the
interpretation of the field $\Psi$ rather than to the evolution
equation~(\ref{MostGeneralEquation})\/.  The form of the initial conditions
could also influence the shape of the stress strain curve considerably, in
particular the onset of instabilities.

Note that the point where the unloading curves meet the main stress-strain
curve moves to higher stresses for increasing plastic deformation,
\textit{i.e.}, the yield surface moves to higher stresses for higher
deformation. In other words, our model equation shows work hardening.
The work hardening is generic to the evolution equation and independent of
the formula for the plastic strain, since the yield surface (\textit{i.e.},
the formation of singularities) moves to higher and higher stresses
in eq.~(\ref{MostGeneralEquation})\/.

The simple Hamilton-Jacobi equation (or anisotropic Burgers equation) (\ref{MostGeneralEquation}) together
with the plastic strain as calculated from eq.~(\ref{StrainLaw}) forms a
theory which generically has all three mysterious features of plasticity,
namely an evolving yield surfaces, cell structure formation (with the
proper cell morphology for the given loading), and work hardening.
However, as already mentioned, there are significant differences
between our model and real plasticity. The cell structure in our model
coarsens while it refines in real systems; also, the cell walls blur upon
unloading in our model. Some of these shortcomings might be related to the
choice of the viscosity solution as weak solution, to the absence of
noise in our evolution equation, or the choice of initial conditions.

The main point we want to make here is, that an evolving yield surface and
work hardening-like behavior is generated generically by the formation of
cusp-singularities in Hamilton-Jacobi equations (and probably other types
of singularities in different evolution equations)\/. The analogy of the singularities in our model with the cell walls is only
qualitative, but it suggests that a mesoscale continuum theory will have 
cell walls as singular structures, that will be central to work hardening
and an evolving yield surface. We hope that the ideas presented in this 
paper will motivate further research in that direction.

\begin{acknowledgments}

Thanks to A.~S.~Argon, P.~R.~Dawson, D.~A.~Hughes, D.~S.~Fisher, J.~Hutchinson, 
C.~R.~Myers, and M.~P.~Miller, for fruitful conversations.
Thanks to G.~Heber and W.~Benger for helping with {\textsc{cactus}} 
on MS-Windows. 
This work was supported by the Cornell Center for Materials Research
NSF (DMR-0079992), by NSF KDI-9873214, NSF ACI-0085969\/, and NSF 9972853,
by Microsoft, and by the Cornell Theory Center. 
\end{acknowledgments}

\end{document}